\pdfoutput=1
\documentclass[
  reprint,
  superscriptaddress,
  amsmath,amssymb,
  prl,
  floatfix
]{revtex4-2}
\usepackage[utf8]{inputenc}
\usepackage[T1]{fontenc}
\usepackage{lmodern}
\usepackage{physics}
\usepackage{caption}
\usepackage{ragged2e}
\usepackage{bbold}
\usepackage{etoolbox}
\usepackage{graphicx}% Include figure files
\usepackage{dcolumn}% Align table columns on decimal point
\usepackage{bm}% bold math
\usepackage[hidelinks]{hyperref}% add hypertext capabilities
\usepackage{amsmath}
\usepackage{amsfonts}
\usepackage{braket}
\usepackage{tcolorbox}

\begin{document}

\preprint{APS/123-QED}

\title{Classical Petermann Factor as a Measure of Quantum Squeezing \\in Photonic Time Crystals}

\author{Younsung Kim}
\thanks{These authors contributed equally to this work.}
\affiliation{Department of Physics, Korea Advanced Institute of Science and Technology, Daejeon 34141, Republic of Korea}

\author{Kyungmin Lee}
\thanks{These authors contributed equally to this work.}
\affiliation{Department of Physics, Korea Advanced Institute of Science and Technology, Daejeon 34141, Republic of Korea}

\author{Changhun Oh}
\affiliation{Department of Physics, Korea Advanced Institute of Science and Technology, Daejeon 34141, Republic of Korea}

\author{Young-Sik Ra}
\affiliation{Department of Physics, Korea Advanced Institute of Science and Technology, Daejeon 34141, Republic of Korea}

\author{Kun Woo Kim}
\email{kunx@cau.ac.kr}
\affiliation{Department of Physics, Chung-Ang University, 06974 Seoul, Republic of Korea}

\author{Bumki Min}
\email{bmin@kaist.ac.kr}
\affiliation{Department of Physics, Korea Advanced Institute of Science and Technology, Daejeon 34141, Republic of Korea}

\begin{abstract}
Photonic time crystals realize a continuum of momentum-resolved $SU(1,1)$ parametric amplifiers. We show that a classical quantity, the Petermann factor of the effective Floquet Bogoliubov–de Gennes (BdG) dynamical matrix, sets the scale of their quantum noise. In stable bands it fixes the Bogoliubov mixing and hence the mean bare-photon occupation of the Floquet vacuum, while in momentum gaps it sets the photon-number prefactor and enhances the squeezing dynamics, with the Floquet growth rate setting the time scale. This converts classical measurements of mode nonorthogonality into quantitative predictions for squeezing and photon generation, and offers a compact design parameter for engineering quantum resources in two-mode BdG platforms.
\end{abstract}

\maketitle

Over the past several decades, two influential yet largely separate narratives have shaped our understanding of quantum noise in optical systems. One traces back to the observation that real lasers can exhibit noise and linewidths far exceeding the Schawlow--Townes prediction~\cite{PhysRev.112.1940}. This discrepancy was explained in large part by recognizing that practical resonators are intrinsically non-Hermitian: their right and left eigenmodes are not orthogonal, and this non-orthogonality enhances the effective spontaneous emission and linewidth by a dimensionless amount now known as the Petermann factor (PF)~\cite{PhysRevA.39.1253,PhysRevA.39.1264, 1070064}. Subsequent experiments on gain-guided and unstable cavities revealed that this factor can become remarkably large, particularly near exceptional points~\cite{Berry01012003}, and modern non-Hermitian photonics views it as a geometric measure of mode sensitivity~\cite{PhysRevResearch.5.033042, PhysRevResearch.6.013044}, linewidth broadening~\cite{Wang2020}, and local density of states (LDOS) enhancement~\cite{Pick:17}. 

A parallel narrative emerged in quantum optics, where the development of two-photon coherent states, $SU(1,1)$ transformations, and optical parametric amplifiers established the concept of squeezed states~\cite{PhysRevA.13.2226, Wodkiewicz:85, PhysRevD.23.1693}. In this framework, vacuum fluctuations are redistributed between conjugate quadratures, and the squeezing parameter quantifies the resulting noise reduction in the squeezed quadrature. It has evolved from a theoretical descriptor of minimum-uncertainty states into a practical resource for precision measurements, ranging from tabletop nonlinear optics experiments to gravitational-wave interferometry~\cite{Andersen_2016,Aasi2013}.

Although both perspectives describe how fluctuations are reshaped and amplified, they have historically been expressed in different languages: the Petermann factor is usually formulated within classical non-Hermitian mode theory, while squeezing belongs to fully quantum, Hermitian dynamics, often formulated in an $SU(1,1)$ framework. Time-varying photonics provides a representative setting in which an underlying Hermitian Hamiltonian structure can give rise to a non-Hermitian dynamical matrix governing field or operator evolution. Within this broader class of time-varying systems, photonic time crystals (PTCs)~\cite{R1_PhysRevA.79.053821,R4_PhysRevB.98.085142,R5_lens.org/077-312-947-644-826,R6_PhysRevB.103.144306,R7_7297162,R8_Park:21,R10_allard2025broadbanddipoleabsorptiondispersive,R11_Asgari:24} provide an ideal platform to unify these narratives. Their momentum gap has been directly observed in microwave circuits and metamaterials~\cite{park_sciadv_doi:10.1126/sciadv.abo6220,wang2023metasurface,lee2025spontaneousemissionlasingphotonic,huang2026observationmomentumbandgapphotonic,6gn2-2v9b}, and an optical realization is being actively pursued~\cite{guo2026plasmonicmetamaterialtimecrystal,galiffi2026coherent}. A PTC implements a continuum of momentum-resolved parametric amplifiers through periodic modulation of a homogeneous medium, producing momentum gaps (MGs) where Floquet quasienergies become complex. Classical analyses reveal that the Petermann factor diverges as one approaches the MG edge, signaling non-Hermitian mode non-orthogonality, while recent quantum studies have associated the MG region with a dynamical Casimir-like effect~\cite{Moore1970,dodonov_physics2010007}, leading to photon-pair creation and squeezing~\cite{doi:10.1021/acsphotonics.4c02293,Bae2025}. As in other contexts, however, previous works on PTCs have treated the Petermann factor and squeezing separately: the Petermann factor of the extended Floquet eigenmodes has been used to explain non-Hermitian reshaping of the momentum-resolved photonic density of states (kDOS) near momentum-gap edges~\cite{5v2w-yg7v,lee2025spontaneousemissionlasingphotonic}, without connecting it to an explicit squeezing description.

In this work, we close this gap by developing a Bogoliubov–de Gennes (BdG) effective-Hamiltonian framework showing that, within the representations adopted here, the classical and quantum monodromy matrices are unitarily equivalent and therefore share identical spectra and Petermann factors. As a result, in the \textit{band regime}, the Petermann factor directly fixes the magnitude of the two-mode Bogoliubov squeezing relating the bare-photon and Floquet quasiparticle vacua, and thereby the associated basis-mismatch occupation. In the \textit{momentum gap}, it sets the photon-number prefactor, while the Floquet growth rate, given by the imaginary part of the quasienergy, sets the time scale of the stroboscopic squeezing dynamics. Classical mode non-orthogonality thus manifests as enhanced quantum squeezing, and the Petermann factor extracted from the BdG dynamical matrix becomes a compact control parameter for squeezing in a broad class of two-mode $SU(1,1)$ bosonic systems, with PTCs providing a particularly transparent and experimentally accessible realization. 

We consider a one-dimensional PTC, a homogeneous medium whose permittivity $\varepsilon(t)$ is modulated periodically in time with period $T=2\pi/\Omega$. The Hamiltonian of a quantum PTC in a specific pair of modes $(k,\,-k)$ can be expressed as~\cite{R3_doi:10.1126/science.abo3324}:
\begin{equation}
    H_k(t)=A(t)(n_k+n_{-k}+1)+B(t) (a_ka_{-k}+a_k^\dagger a_{-k}^\dagger),
    \label{eq:fullH}
\end{equation}
where $n_k=a_k^\dagger a_k$ is the number operator for the $k$-mode. Here we adopt natural units, $\hbar=c=\varepsilon_0=1$. Also, $A(t)=k\bigl[1+\varepsilon^{-1}(t)\bigr]/2$, $B(t)=k\bigl[1-\varepsilon^{-1}(t)\bigr]/2$, where $k$ denotes the magnitude of the wavevector.

Then, it is natural to recast Eq.~(\ref{eq:fullH}) into the BdG form, which leads to:
\begin{equation}
    H_k(t)=\mathbf\Phi^\dagger H_{\mathrm{BdG}}(t)\mathbf\Phi, \quad 
    H_{\mathrm{BdG}}(t)=\begin{bmatrix}A(t) & B(t)\\B(t)& A(t)\end{bmatrix},
    \label{eq:HBdG}
\end{equation}
 where the Nambu basis is defined as $\mathbf\Phi=(a_k, \,a_{-k}^\dagger)^\mathsf{T}$. 
 The operator dynamics is governed by the Heisenberg equation of motion $\partial_t {a}_k=-i[a_k,H_k(t)]$, which in the Nambu basis yields $i\,\partial_t \mathbf\Phi(t) = \mathcal M_q(t)\,\mathbf\Phi(t)$. Here the dynamical matrix is given by $\mathcal M_q(t)\equiv \sigma_z H_{\mathrm{BdG}}(t)$, with the Pauli-$z$ matrix $\sigma_z$. Given the time periodicity $\mathcal{M}_q(t)=\mathcal{M}_q(t+T)$, the stroboscopic behavior of the operators is captured by the monodromy matrix $\mathcal{U}(T,0) = \mathcal{T}\exp\bigl[-i\int_0^T \mathcal{M}_q(t)\,dt\bigr]$, where $\mathcal{T}$ denotes the time-ordering operator. This allows us to define an effective BdG Hamiltonian $H_{\text{BdG}}^{\text{eff}}$ via the relation $\mathcal{U}(T,0)=\exp[-i \sigma_zH_{\text{BdG}}^{\text{eff}}T]$. 
For an explicit representative of \(H_{\mathrm{BdG}}^{\mathrm{eff}}\), we take the matrix logarithm of \( \mathcal U(T,0)\) with the branch chosen such that the real part of the representative quasienergy lies in the Floquet Brillouin zone (FBZ) \((-\pi/T,\pi/T]\). This convention fixes the corresponding matrix representation of \(H_{\mathrm{BdG}}^{\mathrm{eff}}\). At an arbitrary time \(t=nT+\tau\), with \(0\le \tau<T\), the exact propagator factorizes as $\mathcal U(t,0)=\mathcal U(\tau,0)[\mathcal U(T,0)]^n$. While the factor \([\mathcal U(T,0)]^n\) gives the exact period-to-period Nambu basis evolution, \(\mathcal U(\tau,0)\) contains the intra-period micromotion and can be included to recover the full arbitrary-time evolution~\cite{SM}.

\begin{figure*}
    \centering
    \includegraphics[width=\linewidth]{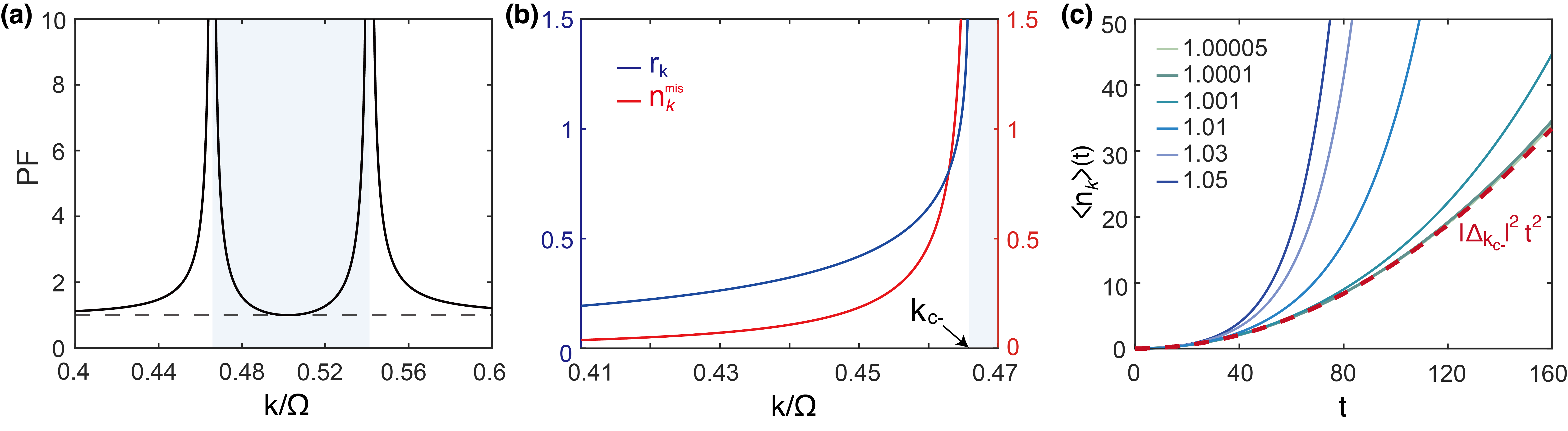}
    \caption{(a) Petermann factor of a PTC with $\varepsilon^{-1}(t)=1-2\alpha\cos{\Omega t}$, where $\alpha=0.15$ and $\Omega=1$. The blue shaded region indicates the MG (unstable) regime, and $K_k$ diverges at the MG edges $k_{c\pm}$. The dashed line represents the lower bound $K_k=1$. (b) Squeezing parameter $r_k$ and the corresponding mismatch occupation $n_k^{\mathrm{mis}}$ in the band (stable) regime. As the momentum approaches the MG edge $k_{c-}$, both quantities increase together with the Petermann factor $K_k$, as quantified by Eqs.~(\ref{eq8}) and (\ref{eq9}). (c) Temporal evolution of the photon number $\braket{\hat{n}_k(t)}$ in the MG, obtained from the effective Floquet Hamiltonian. The curves represent stroboscopic evolution at integer multiples of $T$. Legends indicate the momentum ratio $k/k_{c-}$. As the system approaches the critical point ($k\rightarrow k_{c-}$), the dynamics follow the universal quadratic scaling $|\Delta_{k_{c-}}|^2t^2$, illustrated by the red dashed line.}
    \label{fig:Fig1}
\end{figure*}

Having established the quantum BdG framework, we now clarify its precise correspondence with the classical wave dynamics of the same time-modulated medium. A straightforward calculation shows that the quantum and classical monodromy matrices share the same eigenvalues~\cite{SM}. Consequently, the Heisenberg evolution for the Nambu basis yields the same Floquet quasienergies as the corresponding classical wave dynamics. Since the quantum BdG evolution remains in the $SU(1,1)$ group, its eigenvalues come in reciprocal pairs $\lambda_+\lambda_-=1$ and take the form $\lambda_\pm(k)=e^{\mp i\epsilon_kT}$ in the band (stable) regime and $\lambda_\pm(k)=- e^{\pm {\kappa_k}T}$ in the MG (unstable) regime, where $ {\epsilon_k}$ and $ {\kappa_k}$ are nonnegative real values representing the real quasienergy dispersion and the Floquet growth rate, respectively~\cite{SM}. The resulting quasienergy spectrum, with $ {\epsilon_k}$ and $ {\kappa_k}$ shown explicitly, is illustrated in Fig.~D.2~\cite{SM}. Throughout, we refer to the band (stable) regime as the region $k<k_{c-}$ or $k>k_{c+}$, and to the MG (unstable) regime as $k_{c-}<k<k_{c+}$~\cite{sacha_eswaran2025aspectsquantumgeometryphotonic}.

The full effective Hamiltonian in the stable regime takes the form~\cite{SM}
\begin{equation}
    H_{\mathrm{eff}}=\mathbf\Phi^\dagger H_{\text{BdG}}^{\mathrm{eff}}\mathbf\Phi= {\epsilon_k}(b_{k}^\dagger b_{k}+b_{-k}^\dagger b_{-k}+1),
        \label{eq6}
\end{equation}
where we set $S^{-1}\mathbf\Phi\equiv(b_k,\,b_{-k}^\dagger)^{\mathsf T}$, with $S$ being a Bogoliubov (symplectic) transformation that preserves the bosonic commutation relation $[b_k, b_{k'}^\dagger]=\delta_{kk'}$. With the parametrization $b_k = e^{i\theta}\cosh(r_k)a_k + \sinh(r_k)a^\dagger_{-k}$, the Bogoliubov transformation $S$ diagonalizes $H_{\text{BdG}}^{\text{eff}}$, leading to Eq.~(\ref{eq6}). Thus the quasiparticle band $ {\epsilon_k}$ appearing in the effective Hamiltonian is \emph{exactly} the classical quasienergy band obtained from Maxwell’s equations, establishing a direct classical--quantum correspondence at the level of the Floquet band structure~\cite{SM}.

Denoting $\ket{R_k}$ and $\ket{L_k}$ as the right and left eigenvectors of the effective dynamical matrix respectively, we define the Petermann factor of the one-period monodromy (using the standard Euclidean inner product in Nambu space) to be
\begin{equation}
    K_k = \frac{\braket{L_k|L_k}\braket{R_k|R_k}}
                {|\braket{L_k|R_k}|^2}.
\end{equation}
Due to the relation between the bosonic Nambu basis and the classical bases unitarily related to it~\cite{SM}, $K_k$ simultaneously quantifies the non-orthogonality in both pictures. As shown in Fig.~\ref{fig:Fig1}(a), $K_k$ exhibits a sharp divergence as the momentum approaches the MG edges $k_{c\pm}$. While the classical interpretation links this divergence to non-Hermitian degeneracy, a crucial question remains: how does this classically defined non-Hermitian geometric factor manifest physically in the quantum regime? In what follows, we address this question separately in the band and gap regimes.

We now investigate the Petermann factor in detail by explicitly deriving it from the monodromy matrix. In the band regime $k<k_{c-}$ or $k>k_{c+}$, we have shown that the quasiparticle description allows the Hamiltonian to be written as Eq.~(\ref{eq6}). With this parameterization, the Petermann factor is given by~\cite{SM}
\begin{equation}
    K_{k}=\cosh^2{2r_k}.
    \label{eq8}
\end{equation}
This relation dictates that the Petermann factor is solely governed by the Bogoliubov mixing parameter $r_k$ (or equivalently, the squeezing parameter) that defines the quasiparticle mode $b_k$. 
Figure~\ref{fig:Fig1}(b) shows the momentum dependence of $r_k$ obtained from the stroboscopic time evolution; when viewed together with Fig.~\ref{fig:Fig1}(a), this visually confirms this direct one-to-one correspondence in the band. As the momentum approaches the MG edge, $r_k$ diverges rapidly, representing the extreme limit of the Bogoliubov mixing from the bare mode $a_k$.

The physical significance of this factor is elucidated in Ref.~\cite{5v2w-yg7v,lee2025spontaneousemissionlasingphotonic}, where the weak-coupling spontaneous emission decay rate $\Gamma_D(\omega)$ near the MG edge is analyzed within a classical oscillating-dipole formulation at source frequency $\omega$. In this formulation, the decay rate contains a nonorthogonality contribution associated with the Petermann factor of the extended Floquet eigenmodes (i.e., the left--right overlap in the biorthogonal Green-function expansion). The two-dimensional monodromy Petermann factor considered here provides the corresponding stroboscopic measure of this nonorthogonality enhancement. To provide a quantum interpretation of this mode nonorthogonality, let $\ket{0}_a$ and $\ket{0}_b$ denote the bare photon and Floquet quasiparticle vacua, respectively, defined by $a_{\pm k}\ket{0}_a=0$ and $b_{\pm k}\ket{0}_b=0$. Because these two bases are related by a Bogoliubov transformation, each vacuum is a two-mode squeezed vacuum with respect to the other basis. Accordingly, the same basis mismatch can be expressed as 
\begin{equation}
{}_a\!\bra{0} b_k^\dagger b_k\ket{0}\!_a
=
{}_b\!\bra{0} a_k^\dagger a_k\ket{0}\!_b
=
\frac{\sqrt{K_k}-1}{2}.
\label{eq9}
\end{equation}
Defining this common quantity as the mismatch occupation $n_k^{\mathrm{mis}}$, we obtain the exact relation $\sqrt{K_k}=2n_k^{\mathrm{mis}}+1$. In the band regime, $n_k^{\mathrm{mis}}=\sinh^2 r_k$ is the mean bare-photon occupation of the Floquet vacuum in each mode, equivalently the mean Floquet-quasiparticle occupation of the bare-photon vacuum. These equal vacuum expectation values provide a quantum representation of the left--right eigenmode nonorthogonality of the corresponding classical monodromy. In the Hermitian dynamical-matrix reference limit, the mismatch occupation vanishes, $n_k^{\mathrm{mis}}\to0$, restoring the orthogonal limit $\sqrt{K_k}\to1$.

Having clarified the physical meaning of the Petermann factor in the band regime, we now turn to the MG, where $k_{c-}<k<k_{c+}$ and mode amplification occurs. 
 In this regime, the effective dynamical matrix $\mathcal{M}_q^{\text{eff}}=\sigma_zH_{\text{BdG}}^{\text{eff}}$ takes the form:
\begin{equation}
    \mathcal{M}_q^{\text{eff}}=\begin{pmatrix}
        g_k  & \Delta_k \\ -\Delta_k ^* & -g_k 
    \end{pmatrix}+\mu\mathbb{I},
\end{equation}
where the parameters \(g_k\), \(\Delta_k\), and \(\mu\) parameterize the effective dynamical matrix  \(\mathcal{M}_q^{\text{eff}}\), obtained from the logarithm of the exact monodromy matrix \(\mathcal U(T,0)\). In general, $\mu=0$ or $\pi/T$ in the MG regime; in our case, we have $\mu=\pi/T$. 
The Floquet growth rate is then given by $ {\kappa_k=\sqrt{|\Delta_k|^2-g_k^2}}$, which requires the condition $ {|\Delta_k|>|g_k|}$~\cite{SM}.

With these parameters defined, the Petermann factor in the MG takes the form~\cite{SM}
\begin{equation}
         {K_k = \frac{|\Delta_k|^2}{\kappa_k^2}}.
\end{equation}
This relation illustrates how the Petermann factor manifests itself in the squeezing dynamics. The stroboscopic evolution is captured by a time-dependent squeezing parameter $r_k(t)$, for which one finds the simple relation $\sinh r_k(t)=\sqrt{K_k}\,\sinh( {\kappa_k} t)$. It implies that, for a fixed parametric growth rate $\kappa_k$ (which in PTCs corresponds to the Floquet growth rate), the Petermann factor enhances the initial squeezing rate by $\sqrt{K_k}$~\cite{SM}. We further show that intrinsic loss produces only subleading corrections in a short-time expansion; the Petermann-enhanced initial squeezing rate persists under symmetric Markovian damping~\cite{SM}. Consequently, the average photon number generated from the initial vacuum state, $\braket{\hat{n}_k(t)}\equiv\braket{0|a_k^\dagger(t)a_k(t)|0}$, is given by
\begin{equation}\braket{\hat{n}_k(t)}=K_k\,\sinh^2( {\kappa_k} t). \end{equation} 
Hence, non-orthogonality appears as a multiplicative ``gain factor'' on vacuum amplification.

Even though the growth rate $\kappa_k$ approaches 0 at the MG edge, the diverging Petermann factor compensates for the vanishing growth rate. As $\kappa_k\rightarrow 0$, the photon number exhibits a quadratic scaling, i.e. $\braket{\hat{n}_k(t)}\approx|\Delta_k|^2 t^2$. Since the characteristic time scale $\tau \sim \kappa_k^{-1}$ diverges at the MG edge, this quadratic behavior dominates the dynamics, as illustrated in Fig.~\ref{fig:Fig1}(c). These results can also be understood consistently within the framework of spectral response for a second-order exceptional point~\cite{PhysRevResearch.5.033042, PhysRevResearch.4.023121}.

We now examine the quantum interpretation of the Petermann factor, which measures left--right eigenmode nonorthogonality of the effective dynamical matrix generated by an underlying Hermitian quadratic bosonic Hamiltonian. Since the scalar shift $\mu \mathbb I$ does not affect the eigenvectors or $K_k$, we consider $\widetilde{\mathcal M}_k\equiv \sigma_zH_{\mathrm{BdG}}^{\mathrm{eff}}-\mu\mathbb I$. In the band regime, the $K_k=1$ reference is the Hermitian form $\widetilde{\mathcal M}_k=g_k\sigma_z$, corresponding to $H_{\mathrm{ref}}^{\mathrm{band}}=g_k(a_k^\dagger a_k+a_{-k}^\dagger a_{-k}+1)$. 
In the momentum-gap regime, the \(K_k=1\) reference is anti-Hermitian. Since \(H_{\mathrm{BdG}}^{\mathrm{eff}}\) must remain Hermitian, \(g_k\) is real; anti-Hermiticity of \(\widetilde{\mathcal M}_k\), however, requires \(g_k\) to be purely imaginary, and hence \(g_k=0\). The reference Hamiltonian therefore reduces to $H_{\mathrm{ref}}^{\mathrm{gap}} = \Delta_k^* a_k a_{-k}+\mathrm{h.c.}$. The underlying quadratic Hamiltonian therefore remains Hermitian, while the non-Hermitian structure resides in the dynamical matrix governing operator evolution~\cite{Wakefield2024APLNonHermiticity}. Within the specified basis and stroboscopic convention, $K_k$ is a scale-invariant measure of the basis-relative Bogoliubov mixing relating $\widetilde{\mathcal M}_k$ to these canonical reference forms~\cite{FigueiredoRoque2021,SM}. It therefore quantifies the rotation--squeezing admixture that controls the basis-mismatch occupation in the band and the photon-number prefactor in the momentum gap.

The central message of this work is that a classical nonorthogonality can be read as a quantitative quantum resource; PTCs provide a direct demonstration of this link. Because the classical field evolution and the quantum BdG dynamics are governed by formally equivalent Floquet monodromy matrices, $K_k$ fixes the Bogoliubov mixing in stable bands, while in momentum gaps $K_k$ and the Floquet growth rate $\kappa_k$ jointly control the time-dependent photon generation. Since $K_k$ and $\kappa_k$ are fixed by the one-period monodromy~\cite{SM}, the corresponding squeezing and photon yield follow from the algebraic relations derived here; the corresponding Floquet-mode nonorthogonality also governs classical signatures such as decay-rate enhancement and kDOS reshaping~\cite{5v2w-yg7v,lee2025spontaneousemissionlasingphotonic}. More broadly, this identifies biorthogonal mode geometry as a practical design knob for parametric quantum noise in a variety of quadratic bosonic platforms admitting an effective two-mode $SU(1,1)$ BdG reduction~\cite{A1_PhysRevLett.103.147003,A2_Wilson2011,Murch2013,Hudelist2014}. In short, biorthogonality is not a nuisance here; it is the resource that sets the quantum noise and the resulting squeezing and photon yield.

\begin{acknowledgments}
We thank J. B. Pendry, P. A. Huidobro and  T. F. Allard for insightful discussions. This work is supported by the National Research Foundation of Korea (NRF) through the government of Korea (RS-2022-NR070636) and the Samsung Science and Technology Foundation (SSTF-BA240202). K.W.K. acknowledges financial support from the Basic Science Research Program through the NRF funded by the Ministry of Education (no. RS-2025-00521598) and the Korean Government (MSIT) (no. 2020R1A5A1016518). Y.-S.R. acknowledges the support from Institute of Information \& Communications Technology Planning \& Evaluation (IITP) grant (RS-2025-25464959). C.O. was supported by the NRF grants funded by the MSIT (No. RS-2024-00431768 and No. RS-2025-00515456) and the IITP grants funded by the MSIT (No. RS-2025-02283189, No. RS-2025-02263264, and No. RS-2025-25464990). K.L. also acknowledges support from the KAIST Jang Young Sil Fellow Program.    
\end{acknowledgments}

\makeatletter\global\let\@FMN@list\@empty\makeatother

\clearpage
\onecolumngrid
\setcounter{page}{1}
\setcounter{equation}{0}
\setcounter{section}{0}
\setcounter{figure}{0}
\setcounter{table}{0}
\renewcommand{\theequation}{S\arabic{equation}}
\renewcommand{\thefigure}{S\arabic{figure}}
\renewcommand{\thetable}{S\arabic{table}}
\begingroup
\hypersetup{pageanchor=false}
\begin{center}
{\large\bfseries Supplemental Material: Classical Petermann Factor as a Measure of\\ Quantum Squeezing in Photonic Time Crystals\par}
\vspace{0.75em}
Younsung Kim\textsuperscript{1,*}, Kyungmin Lee\textsuperscript{1,*}, Changhun Oh\textsuperscript{1}, Young-Sik Ra\textsuperscript{1},\\
Kun Woo Kim\textsuperscript{2,$\dagger$}, and Bumki Min\textsuperscript{1,$\ddagger$}\\[0.45em]
\textsuperscript{1}\textit{Department of Physics, Korea Advanced Institute of Science and Technology, Daejeon 34141, Republic of Korea}\\
\textsuperscript{2}\textit{Department of Physics, Chung-Ang University, 06974 Seoul, Republic of Korea}
\end{center}
\hypersetup{pageanchor=true}
\endgroup
\begin{center}
\textbf{CONTENTS}
\end{center}
\begingroup
\setlength{\parindent}{0pt}
A. Classical-to-quantum correspondence \dotfill \pageref{smsec:A}\par
B. $\mathfrak{su}(1,1)$ algebra and evolution \dotfill \pageref{smsec:B}\par
C. Stable vs unstable regime in $SU(1,1)$ \dotfill \pageref{smsec:C}\par
D. Effective Hamiltonian and band correspondence \dotfill \pageref{smsec:D}\par
E. Petermann factor in quantum formalism \dotfill \pageref{smsec:E}\par
F. Petermann factor as a squeezing measure in the unstable regime \dotfill \pageref{smsec:F}\par
G. Effect of loss on the Petermann-squeezing relation \dotfill \pageref{smsec:G}\par
References \dotfill \pageref{smsec:refs}\par
\endgroup

\section{A. Classical-to-quantum correspondence}\label{smsec:A}

We consider a lossless electrodynamical system whose dynamics is governed by a Hamiltonian that is quadratic in the canonical fields (we set $\hbar=c=1$). Starting from Maxwell's equations, which are first-order differential equations, one can introduce canonical variables such that the basis elements satisfy the Poisson bracket relation. In this canonical basis the dynamics takes the Nambu form
\begin{equation}
i\,\partial_t
\begin{pmatrix}
\alpha_k\\
\alpha_{-k}^*
\end{pmatrix}
=
\mathcal M
\begin{pmatrix}
\alpha_k\\
\alpha_{-k}^*
\end{pmatrix},
\end{equation}
with \(
\{\alpha_k,\alpha_{k'}^*\}=-i\delta_{kk'}.
\) Then, the direct quantization process $i\{\;\;,\;\;\} \rightarrow [\;\;,\;\;] $ leads to the equation

\begin{equation}
    i\,\partial_t  \begin{pmatrix}
        \hat a_k \\ \hat a_{-k}^\dagger
    \end{pmatrix} = \mathcal M\begin{pmatrix}
        \hat a_k \\ \hat a_{-k}^\dagger
    \end{pmatrix},
\end{equation}
where $\mathcal M=\sigma_zH_{\text{BdG}}$, with $H_{\text{BdG}}$ Hermitian. 
This indicates that the classical and quantum dynamics share the same dynamical matrix, and therefore share its spectrum $\{\omega_n\}$. This formalism can be naturally generalized to multimode systems, as the classical and quantum correspondence remains valid in general symplectic groups $Sp(2n, \mathbb{R})$~\cite{S-Wunsche2000Symplectic}. Moreover, $\mathcal M$ is $\sigma_z$-pseudo-Hermitian, i.e.,
\begin{equation}
\mathcal M^\dagger \sigma_z=\sigma_z\mathcal M.
\end{equation} 
So its eigenvalues $\{\omega_n\}$ and corresponding eigenvectors $\{v_n\}$ satisfy
\begin{equation}
(\omega_m^*-\omega_n)\,v_m^\dagger\sigma_z v_n=0.
\end{equation}
That is, orthogonality is naturally defined with respect to the Krein inner product
$\langle v_m,v_n\rangle_K:=v_m^\dagger \sigma_z v_n$ rather than the Euclidean product $v_m^\dagger v_n$.
In the case of a two-dimensional matrix, for distinct real eigenvalues
$\omega_1\neq \omega_2\in\mathbb R$, the eigenmodes satisfy $\langle v_1,v_2\rangle_K=0$.
If the eigenvalues form a non-real complex-conjugate pair, $\omega_2=\omega_1^*\notin\mathbb R$,
then $\langle v_1,v_1\rangle_K=0$ and $\langle v_2,v_2\rangle_K=0$.

A particularly relevant example is the quantization of a photonic time crystal (PTC), whose permittivity $\varepsilon(t)$ is periodically modulated in time. In the explicit PTC case below, we restore the physical constants $\hbar$ and $c$, which were set to unity in the general discussion above. Starting from Maxwell's equations and performing a spatial Fourier transform, we can write the equations in the following form:
\begin{align}
    i\partial_t \begin{bmatrix}
        \tilde{\mathbf{D}}(\mathbf{k},t)\\\tilde{\mathbf{H}}(\mathbf{k},t)\end{bmatrix}=\begin{bmatrix}
            \mathbb{0}_3 & -\mathbf{k}\times \\ \frac{1}{\varepsilon \mu_0}\mathbf{k}\times &\mathbb{0}_3 
        \end{bmatrix}\begin{bmatrix}
        \tilde{\mathbf{D}}(\mathbf{k},t)\\\tilde{\mathbf{H}}(\mathbf{k},t)\end{bmatrix},
\end{align}
where $\tilde{\mathbf{D}}$ and $\tilde{\mathbf{H}}$ denote the three-component Fourier amplitudes of the displacement and magnetic fields, respectively. Now, we specialize to the one-dimensional case with propagation along the $z$ direction, $\mathbf{k}=k\hat{\mathbf e}_z$. Then, one can decompose the dynamics into independent polarization sectors. Restricting ourselves to the $(\tilde D_x,\, \tilde H_y)$ sector, we obtain
\begin{align}
    i\partial_t\begin{bmatrix}
        \tilde D_x(k,t) \\ \tilde H_y(k,t)
    \end{bmatrix}=\begin{bmatrix}
        0 &  k \\ \frac{k}{\varepsilon\mu_0} & 0
    \end{bmatrix}\begin{bmatrix}
        \tilde D_x(k,t) \\ \tilde H_y(k,t)
    \end{bmatrix}.
\end{align}
To connect this Maxwell dynamics to the Nambu basis $(a_k,\,a_{-k}^\dagger)^\mathsf{T}$ representation, it is natural to introduce the reference energy-normalized field vector
\begin{equation}
    \mathbf{F}_k=\begin{bmatrix}
        \tilde D_x/\sqrt{\varepsilon_{\text{ref}}},\, \sqrt{\mu_0}\tilde H_y
    \end{bmatrix}^\mathsf{T},
\end{equation}
where $\varepsilon_{\text{ref}}$ denotes a fixed reference permittivity, taken to be $\varepsilon_0$ in the main text. The dynamics can then be recast as
\begin{align}
    i\partial_t\mathbf{F}_k(t)
    &=
    \mathsf{M}_{F,k}(t)\mathbf{F}_k(t),\label{eq:classdy}\\
    \mathsf{M}_{F,k}(t)
    &=
    \omega_{\mathrm{ref},k}
    \begin{bmatrix}
        0 & 1\\
        f(t) & 0
    \end{bmatrix},\label{eq:classdymat}
\end{align}
where
\begin{equation}
    \omega_{\mathrm{ref},k}
    \equiv
    \frac{k}{\sqrt{\varepsilon_{\mathrm{ref}}\mu_0}},
    \qquad
    f(t)
    \equiv
    \frac{\varepsilon_{\mathrm{ref}}}{\varepsilon(t)}.\label{eq:referenceperm}
\end{equation}
For real $\varepsilon(t)$, the Maxwell generator satisfies
\begin{equation}
    \mathsf{M}_{F,k}^{\dagger}(t)\sigma_x
    =
    \sigma_x\mathsf{M}_{F,k}(t),
\end{equation}
and is therefore $\sigma_x$-pseudo-Hermitian.
In the static limit where permittivity is $\varepsilon_{\mathrm{ref}}$,
$f=1$, so that
$\mathsf{M}_{F,k}=\omega_{\mathrm{ref},k}\sigma_x$
is manifestly Hermitian and
$\mathbf F_k^{\dagger}\mathbf F_k$ is proportional to the conserved electromagnetic energy.

Furthermore, the energy contribution associated with the selected \(k\) and polarization sector in the general time-varying setting can be expressed directly
in terms of the Maxwell generator as
\begin{align}
    \mathcal H_k^{\mathrm{cl}}(t)
    &=
    \frac{1}{2\omega_{\mathrm{ref},k}}
    \mathbf F_k^\dagger(t)
    \sigma_x\mathsf M_{F,k}(t)
    \mathbf F_k(t) \nonumber\\
    &=
    \frac12
    \mathbf F_k^\dagger(t)
    \begin{bmatrix}
        f(t) & 0\\
        0 & 1
    \end{bmatrix}
    \mathbf F_k(t) \nonumber\\
    &=
    \frac12
    \left[
        \frac{|\tilde D_x(k,t)|^2}{\varepsilon(t)}
        +
        \mu_0|\tilde H_y(k,t)|^2
    \right].
\end{align}

We now follow the canonical quantization procedure of Lyubarov \textit{et al.}~\cite{S-R3_doi:10.1126/science.abo3324}, applied here to the reduced two-component Maxwell equation in Eq.~\eqref{eq:classdy}.  Below, we briefly summarize the procedure and make explicit the connection between the energy-normalized Maxwell basis and the corresponding quantum dynamics in the bosonic Nambu representation. In Coulomb gauge with no free sources, the Lagrangian density of the field is given by 
\begin{equation}
    \mathcal{L}=\frac{\varepsilon(t)}{2}(\partial_t\mathbf{A})^2-\frac{1}{2\mu_0}(\nabla\times \mathbf{A})^2,
\end{equation}
and the canonical conjugate field is
\begin{equation}
    \mathbf\Pi=\frac{\partial\mathcal{L}}{\partial \dot{\mathbf{A}}}=-\varepsilon(t)\mathbf{E}=-\mathbf{D}.
\end{equation}
Restricting the dynamics to the one-dimensional transverse sector introduced above, the canonical fields can be expanded as
\begin{align}
   A_x(z,t)=& \frac{1}{\sqrt{\varepsilon_0}}
\int\frac{dk}{\sqrt{2\pi}}\,
e^{ikz} q_k(t),\\
\Pi_x(z,t)
=&
\sqrt{\varepsilon_0}
\int\frac{dk}{\sqrt{2\pi}}\,
e^{-ikz} p_k(t).
\end{align}
The canonical fields satisfy the equal-time Poisson bracket $\{A_x(z,t),\Pi_x(z',t)\}
=
\delta(z-z')$, and under the Fourier convention above, the momentum-space amplitudes satisfy the Poisson bracket relation $\{q_k(t), \,p_{k'}(t)\}=\delta(k-k')$.

Then, using the relations $\mathbf{D}=-\mathbf{\Pi}$ and $\mathbf{H}=(\nabla\times\mathbf{A})/\mu_0$, we arrive at the relation
\begin{align}
    \tilde D_x(k,t)=
-\sqrt{\varepsilon_0}\,p_{-k}(t),\quad
\tilde H_y(k,t)
=\frac{ik}{\mu_0\sqrt{\varepsilon_0}}\,q_k(t). \label{eq:DH}
\end{align}
Substituting these relations into the $k$-resolved electromagnetic
energy derived above, we obtain
\begin{equation}
    \mathcal H_k^{\mathrm{cl}}(t)
    =
    \frac12
    \left[
        \frac{\varepsilon_0}{\varepsilon(t)}
        p_{k}p_{k}^*
        +
        c^2k^2q_k q_k^*
    \right],
\end{equation}
where the reality of the fields implies $q_k^*=q_{-k}$ and $p_k^*=p_{-k}$.

This time-dependent harmonic oscillator form motivates the introduction of classical complex canonical amplitudes associated with the fixed reference permittivity and the corresponding reference frequency defined in Eq.~\eqref{eq:referenceperm}:
\begin{align}
    \alpha_k
    &=
    \sqrt{\frac{ckn_{\mathrm{ref}}}{2\hbar}}\,
    q_k
    +
    i
    \sqrt{\frac{1}{2\hbar ckn_{\mathrm{ref}}}}\,
    p_{-k},
    \\
    \alpha_{-k}^{*}
    &=
    \sqrt{\frac{ckn_{\mathrm{ref}}}{2\hbar}}\,
    q_k
    -
    i
    \sqrt{\frac{1}{2\hbar ckn_{\mathrm{ref}}}}\,
    p_{-k},
\end{align}
where $n_{\mathrm{ref}}^2\equiv \varepsilon_{\text{ref}}/\varepsilon_0$ and $k>0$. Then with Eq.~(\ref{eq:DH}), we finally arrive at the relation
\begin{align}
    &\mathbf{F}_k = i\sqrt{\hbar \omega_{\text{ref},k}}\mathsf{U}\bm{\alpha}_k,\\
     \mathsf{U}=\dfrac{1}{\sqrt{2}}&\begin{bmatrix}
        1 & -1 \\ 1 & 1 
    \end{bmatrix},\quad\bm{\alpha}_k=\begin{bmatrix}
        \alpha_k \\ \alpha^*_{-k}
    \end{bmatrix},
\end{align}
and Eq.~(\ref{eq:classdy}) can be written as
\begin{equation}
    i\partial_t\bm{\alpha}_k=\mathsf{U}^\dagger\mathsf{M}_{F,k}(t)\mathsf{U}\bm{\alpha}_k.
\end{equation}
Promoting the amplitudes $\alpha_k$ to the quantized ladder operator $\hat a_k$, and explicitly writing the components of the dynamical matrix, one arrives at 
\begin{align}
     i\hbar \partial_t\begin{bmatrix}
        \hat a_k\\ \hat a_{-k}^\dagger
    \end{bmatrix}=&\,\frac{\hbar c k}{n_\text{ref}}\frac{1}{2}\begin{bmatrix}1+f(t) &  1-f(t) \\ -(1-f(t))  & -(1+f(t)) \end{bmatrix}\begin{bmatrix}
        \hat a_k\\ \hat a_{-k}^\dagger 
    \end{bmatrix}.
\end{align}
Here, under the unitary basis transformation $\mathsf{U}$, the $\sigma_x$-pseudo-Hermitian structure of the Maxwell generator is mapped to the conventional $\sigma_z$-pseudo-Hermitian structure of the bosonic BdG dynamical matrix, since
\begin{equation}
 \mathsf U^\dagger\sigma_x\mathsf U = \sigma_z.
\end{equation}

The same transformation also maps the classical electromagnetic energy contribution associated with the $\pm k$ sector. Indeed, defining
\begin{equation}
    \mathsf M_{N,k}(t)
    \equiv
    \mathsf U^\dagger
    \mathsf M_{F,k}(t)
    \mathsf U,
\end{equation}
and combining the contributions from $k$ and $-k$, we obtain
\begin{equation}
    \mathcal H_{(k,-k)}^{\mathrm{cl}}(t)
    =
    \hbar
    \boldsymbol\alpha_k^\dagger
    \sigma_z\mathsf M_{N,k}(t)
    \boldsymbol\alpha_k.
\end{equation}
Here,
\begin{equation}
    \hbar\sigma_z\mathsf M_{N,k}(t)
    =
    \frac{\hbar \omega_{\mathrm{ref},k}}{2}
    \begin{bmatrix}
        1+f(t) & 1-f(t)\\
        1-f(t) & 1+f(t)
    \end{bmatrix}
\end{equation}
is Hermitian and therefore constitutes the corresponding bosonic
BdG Hamiltonian $H_{\text{BdG}}(t)$ after quantization.

In the main text, $k>0$ labels the counterpropagating pair $(k,-k)$, and the corresponding pair Hamiltonian is denoted by $\hat H_k(t)$:
\begin{align}
    \hat H_{k}(t)
    =&
    \hbar
    \hat{\boldsymbol\Phi}_k^\dagger
    \sigma_z\mathsf M_{N,k}(t)
    \hat{\boldsymbol\Phi}_k\\
    \equiv & \hat{\boldsymbol\Phi}_k^\dagger H_{\text{BdG}}(t)\hat{\boldsymbol\Phi}_k,
    \qquad
    \hat{\boldsymbol\Phi}_k
    = 
    \begin{bmatrix}
        \hat a_k\\
        \hat a_{-k}^\dagger
    \end{bmatrix}.
\end{align}
For notational convenience, we suppress the operator hats hereafter, as well as in the main text.

For a PTC, however, the instantaneous eigenvalues and eigenvectors of
the dynamical matrix do not in general define the Floquet modes of the
periodically driven system. Instead, according to Floquet theory, the
relevant modes and quasifrequencies are obtained from the eigenvalue
problem of the one-period evolution operator, or monodromy matrix,
\begin{equation}
    \mathcal U(T,0)
    =
    \mathcal T
    \exp\left[
        -i\int_0^T dt\,\mathsf M(t)
    \right].
\end{equation}
Since $\mathsf U$ is time independent, the Maxwell and Nambu
monodromy matrices are related by the same unitary transformation,
\begin{equation}
    \mathcal U_{N,k}(T,0)
    =
    \mathsf U^\dagger
    \mathcal U_{F,k}(T,0)
    \mathsf U,
\end{equation} 
and therefore possess the same Floquet spectrum.

Moreover, the corresponding Petermann factor is unchanged under this unitary transformation. Let $\ket{R_{k}^{F}}$ and $\bra{L_{k}^{F}}$ denote the right and left eigenvectors of $\mathcal U_{F,k}(T,0)$. Their counterparts in the Nambu representation are
\begin{equation}
    \ket{R_{k}^{N}}
    =
    \mathsf U^\dagger\ket{R_{k}^{F}},
    \qquad
    \bra{L_{k}^{N}}
    =
    \bra{L_{k}^{F}}\mathsf U.
\end{equation}
The Petermann factor,
\begin{equation}
    K_k
    =
    \frac{
        \braket{R_k|R_k}
        \braket{L_k|L_k}
    }{
        \left|\braket{L_k|R_k}\right|^2
    },
\end{equation}
is therefore identical in the two representations,
\begin{equation}
    K_k^{N}
    =
    K_k^{F},
\end{equation}
because the norms and biorthogonal overlap are preserved by
$\mathsf U^\dagger\mathsf U=\mathbb I$.

\section{B. $\mathfrak{su}(1,1)$ algebra and evolution}\label{smsec:B}

For any time interval, the evolution matrix is given as 
\begin{equation}
    \mathcal{U}(t_2,t_1)=\mathcal{T}\exp\big[-i\int_{t_1}^{t_2} \sigma_zH_{\text{BdG}}(t)dt  \big].
\end{equation}\newline
Here, the dynamical matrix is explicitly
\begin{align}
    \sigma_zH_{\text{BdG}}(t)=\begin{bmatrix}A(t) & B(t)\\ -B(t) & -A(t) 
    \end{bmatrix}.
\end{align}

\noindent Meanwhile, the $\mathfrak{su}(1,1)$ algebra can be defined with a basis 
\begin{equation}
    K_x=\frac{\sigma_x}{2},\quad K_y=\frac{\sigma_y}{2},\quad K_z=i\frac{\sigma_z}{2}
\end{equation} such that
\begin{equation}
    [K_x,K_y]=K_z,\quad[K_y,K_z]=-K_x,\quad[K_z,K_x]=-K_y
\end{equation}
The generator of the evolution can be written as $-i\sigma_z H_{\mathrm{BdG}}(t)=2\big[B(t)K_y-A(t)K_z\big]$ with real $A(t)$ and $B(t)$. Therefore, although the time-evolution operator involves a time-ordered exponential, the generator remains within the $\mathfrak{su}(1,1)$ Lie algebra at all times, and the resulting evolution operator stays in the associated $SU(1,1)$ group (up to an overall phase if an identity component is included).

\section{C. Stable vs unstable regime in $SU(1,1)$}\label{smsec:C}
If the evolution operator $\mathcal{U}$ is an element of $SU(1,1)$, we could parameterize it as follows:
\begin{equation}
    \mathcal{U}=\begin{bmatrix}
        a & b \\ b^* & a^*
    \end{bmatrix},\quad|a|^2-|b|^2=1,
\end{equation}
where $a$ and $b$ are complex numbers. Here, its characteristic polynomial is given as:
\begin{equation}
    P(\lambda)=\lambda^2-\text{Tr}(\mathcal{U})\lambda +1.
\end{equation}
Hence, letting two eigenvalues be $\lambda_\pm$, $\lambda_+ \lambda_-=1$ holds. Imposing the trace condition, we could separate it into three cases:
\begin{equation}
    \begin{cases}
    \mathrm{(i)}\;|\text{Tr}(\mathcal{U})|<2\quad&\text{(Elliptic)} \\
    \mathrm{(ii)}\;|\text{Tr}(\mathcal{U})|=2\quad&\text{(Parabolic)} \\
    \mathrm{(iii)}\;|\text{Tr}(\mathcal{U})|>2\quad&\text{(Hyperbolic)}
\end{cases}
\label{eq:trace}
\end{equation}
Solving the characteristic polynomial under the trace conditions in Eq.~(\ref{eq:trace}), we obtain the following classifications of eigenvalues. In case (i), the eigenvalues form a complex-conjugate pair, and we could let $\lambda_\pm=e^{\pm i\theta}$, where $\theta\in \mathbb{R}$ defined modulo $2\pi$. This corresponds to the stable regime. In case (ii), the eigenvalues are degenerate, $\lambda=\pm1$; generically the matrix becomes non-diagonalizable, except for the trivial case $\mathcal U=\pm \mathbb{I}$. And in case (iii), the eigenvalue pair is purely real, and we could let it be $\lambda$ and $1/\lambda$. It corresponds to the unstable regime.

\section{D. Effective Hamiltonian and band correspondence}\label{smsec:D}
Here, we derive the explicit expression for the stroboscopic Hamiltonian for both stable and unstable regimes. In particular, by relating the Heisenberg dynamics in the Nambu basis to the effective Hamiltonian written in the Fock state basis, we show that the quasienergy spectrum emerges in a Bogoliubov spectrum-like form. First, briefly recall some basic facts about Floquet systems. The time periodicity of a dynamical matrix $\mathcal{M}(t)=\mathcal{M}(t+T)$ makes it possible to capture the stroboscopic behavior via the monodromy matrix $\mathcal{U}(T,0)=\mathcal{T}\exp(-i\int_0^T\mathcal{M}(t)dt)$, where $\mathcal{T}$ denotes the time-ordering operator. Here, the quasienergies $\{\nu_i\}$ are obtained from the eigenvalues of the monodromy matrix $\lambda_i =e^{-i\nu_iT}$ and are defined modulo $2\pi/T$.

The stroboscopic formulation can be connected to the full arbitrary-time dynamics in a direct way. The evolution operator at arbitrary time is governed by the time-ordered evolution operator
\begin{equation}
    \mathcal{U}(t,0)=\mathcal{T}\exp\bigl[-i\int_0^t \mathcal{M}(t')\,dt'\bigr].
\end{equation}

For \(t=nT+\tau\), with \(0\le \tau<T\), it factorizes as $\mathcal U(t,0) =\mathcal U(\tau,0)[\mathcal{U}(T,0)]^n$, directly showing \([\mathcal{U}(T,0)]^n\) gives the exact period-to-period evolution, while \(\mathcal U(\tau,0)\) contains the intra-period micromotion. 
At stroboscopic times, \(\tau=0\), this reduces to $\mathcal U(nT,0)=[\mathcal{U}(T,0)]^n$.

\begin{figure}[!t]
    \centering
\includegraphics[width=\linewidth]{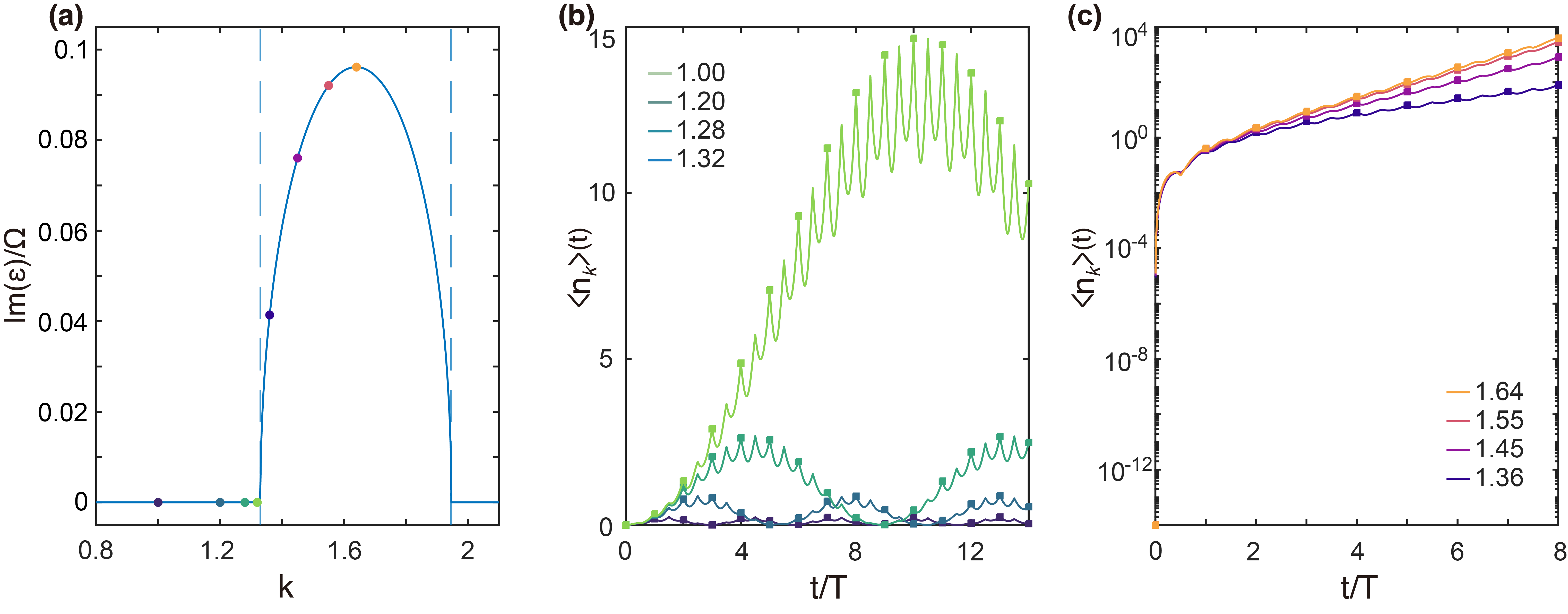}
    \caption*{\justifying FIG. D.1. Stroboscopic dynamics versus full time evolution in a two-step photonic time crystal. The inverse permittivity is modulated as a two-step function \(\varepsilon^{-1}=1.6\) and \(\,0.4\), with equal step durations and period \(T=2\). We plot the photon population generated from the vacuum, $\braket{\hat{n}_k(t)}\equiv\braket{0|a_k^\dagger(t)a_k(t)|0}$. \textbf{(a)} Imaginary part of the Floquet band structure as a function of wavenumber \(k\). The dashed lines mark the momentum-gap edges, and the colored markers indicate the modes we select. \textbf{(b)} Pass-band modes, where the population remains bounded. \textbf{(c)} Momentum-gap modes on a logarithmic scale, where the population is exponentially amplified. Solid curves show the full arbitrary-time evolution including intra-period micromotion, while square markers show the exact stroboscopic values at \(t=nT\), computed from \([\mathcal U(T,0)]^n\). Legends indicate the wavenumber \(k\).}
    \label{fig:full_strob}
\end{figure}

The effective dynamical matrix $\sigma_z H_{\mathrm{BdG}}^{\mathrm{eff}}$ used in our work is then defined by $\mathcal{U}(T,0)=\exp\left[-i\sigma_zH_{\mathrm{BdG}}^{\mathrm{eff}}T\right]$.  For $t\neq nT$, the residual factor $\mathcal{U}(\tau,0)$ generates additional intra-period dynamics, and retaining it yields the full arbitrary-time evolution. Fig.~D.1 illustrates this point in the simplest two-step PTC setting: the solid curves in Fig.~D.1(b,c) show the full evolution including micromotion, while the markers track the exact stroboscopic dynamics $[\mathcal{U}(T,0)]^n$ that our analysis is built on.

We start from the observation that the monodromy matrix $\mathcal{U}_F\equiv \mathcal{U}(T,0)$ belongs to the $SU(1,1)$ group. In a stable regime, $\mathcal U_F$ and $\text{diag}(e^{- iT\epsilon_k},\,e^{ iT\epsilon_k})$ are in the same conjugacy class~\cite{S-Inglima:2018oda}. That is, there exists $S\in SU(1,1)$ such that 
\begin{equation}
    \mathcal U_F=S\,\begin{pmatrix}
        e^{- iT\epsilon_k} & 0 \\ 0&e^{ iT\epsilon_k} 
    \end{pmatrix}\,S^{-1}.
\end{equation}

We now take a matrix logarithm of $\mathcal U_F$ and choose its branch such that the representative quasienergy lies in the Floquet Brillouin zone (FBZ) $(-\pi/T,\pi/T]$, which uniquely determines the quasienergy $\epsilon_k$ within this interval. The logarithm is then given by
\begin{align}
    \log{\mathcal U_F}=-iT\bigg[S\begin{pmatrix}
        \epsilon_k & 0 \\ 0 & -\epsilon_k
    \end{pmatrix}S^{-1}\bigg].
\end{align}
Here, we define $\tilde{D}=\text{diag}(\epsilon_k, -\epsilon_k)$, which leads to the relation $S\tilde{D}S^{-1}=\sigma_z H_{\text{BdG}}^{\text{eff}}$ via $\mathcal{U}_F=e^{-iT\sigma_z H_{\text{BdG}}^{\mathrm{eff}}}.$ Now, we are ready to explicitly derive the form of $H_\text{eff}$. We can write the following relation:
\begin{equation}
    \begin{aligned}H_{\mathrm{eff}}&=\mathbf\Phi^\dagger H_{\text{BdG}}^{\mathrm{eff}}\mathbf\Phi\\
&=\mathbf\Psi^\dagger(S^\dagger H_{\text{BdG}}^{\mathrm{eff}}S)\mathbf\Psi\\
&=\mathbf\Psi^\dagger(S^\dagger \sigma_z\cdot S\tilde D S^{-1}\cdot S)\mathbf\Psi\\
&=\mathbf\Psi^\dagger\underbrace{(S^\dagger \sigma_zS)}_{=\sigma_z\,\,\,\because S\in SU(1,1)}\tilde D\mathbf\Psi\\
&=\mathbf\Psi^\dagger\sigma_z\tilde D\mathbf\Psi\\
&=\mathbf\Psi^\dagger\begin{pmatrix} \epsilon_k & \; \\ \; &\epsilon_k\end{pmatrix}\mathbf\Psi\\
&=\epsilon_k(b_{k}^\dagger b_{k}+b_{-k}^\dagger b_{-k}+1).
\end{aligned}
\end{equation}
Here we set $\mathbf\Psi\equiv S^{-1}\mathbf\Phi=(b_k,\,b_{-k}^\dagger)^{\mathsf T}$, where $S$ is a symplectic transformation that preserves the bosonic commutation relation $[b_k, b_{k'}^\dagger]=\delta_{kk'}$. We have used the relation that the symplectic structure preserves the relation $S^\dagger \sigma_z S=\sigma_z$. Also, we should note that the Fock-state quasienergy spectrum shown in Fig.~D.2(b) is calculated \textit{before} FBZ folding. This representation naturally arises when we take the matrix logarithm of the monodromy operator to construct the effective Hamiltonian.

\begin{figure*}
    \centering
    \includegraphics[width=0.7\linewidth]{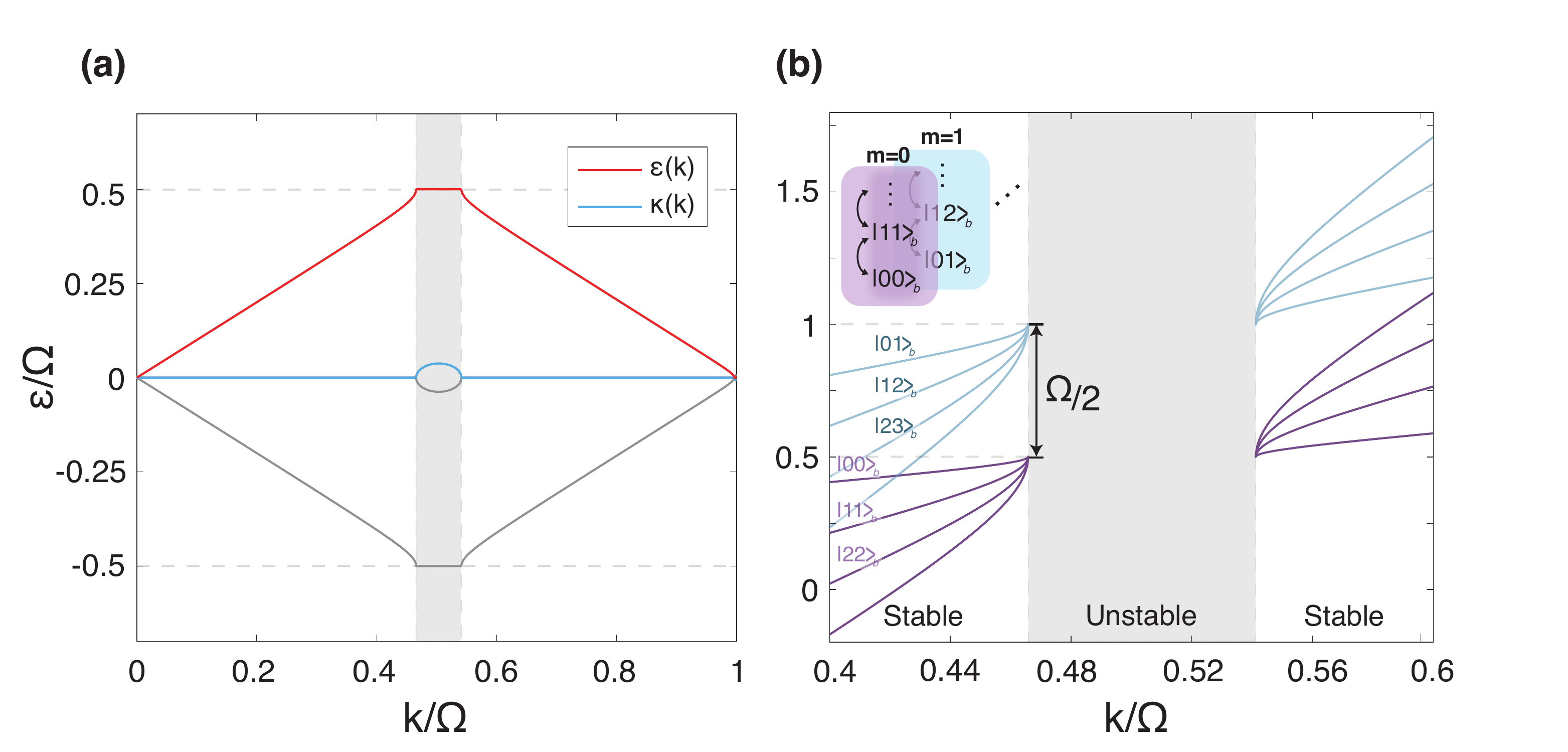}
    \caption*{\justifying FIG. D.2. (a) Floquet band structure inside Floquet Brillouin zone (FBZ) from quantum BdG dynamical matrix. The gray-shaded region indicates the momentum gap (MG), and $\epsilon_k $ and $\kappa_k$ indicate the real and imaginary part of the quasienergy, respectively. (b) Spectrum obtained from the effective Hamiltonian $H_{\text{eff}}$. The gray-shaded region indicates the MG, while blue and purple lines denote the spectrum for the momentum sectors $q=1$ and $q=0$, respectively. The discrete branches correspond to the Fock-state energy levels $(n_k+m_{-k}+1)\epsilon_k$.}
    \label{fig:placeholder}
\end{figure*}

Now, we consider the unstable case, where the quasienergies become complex-valued. In this regime, the eigenvalues of $\mathcal{U}_F$  are real and reciprocal, i.e., $(\lambda,\,\lambda^{-1})$. Writing $\lambda=e^{-i\nu T}$, this implies that the quasienergies are given as $(\nu,\,-\nu)$ modulo $2\pi/T$.
Since $\lambda, \,\lambda^{-1}\in \mathbb{R}$, the real part of $\nu$ (denoted by $\epsilon_k$) must satisfy $e^{-i\epsilon_k T}=\pm 1$. Equivalently, denoting this value by $\mu\equiv \epsilon_k$ for brevity, $\mu=0$ or $\pi/T$ within the FBZ $(-\pi/T,\pi/T]$.

In Fig.~D.2(a), the gray-shaded momentum gap region corresponds to $\mu=\pi/T$, yielding negative real eigenvalues. Consequently, the eigenvalue pair becomes $\lambda_\pm=-e^{\pm\kappa_k T}$. That is, $\text{Tr}\,\mathcal{U}_F=-2\cosh(\kappa_k T)<-2$.

\noindent Here as well, we note that the monodromy matrix $\mathcal{U}_F\in SU(1,1)$ having eigenvalue pair $\lambda_\pm=e^{\pm\kappa_kT-i\mu T}$ is conjugate to the following canonical hyperbolic representative within the group~\cite{S-Inglima:2018oda}:
\begin{equation}
    e^{-i\mu T\mathbb{I}}\begin{pmatrix}\cosh\kappa_k{T}&\sinh\kappa_k{T} \\ \sinh\kappa_k{T}& \cosh\kappa_k{T}\end{pmatrix}=e^{\kappa_k{T} \sigma_x-i\mu T\mathbb{I}}.
    \label{eq:hyp}
\end{equation}
Moreover, the hyperbolic elements are not conjugate to any diagonal matrix in $SU(1,1)$, since diagonal matrices in $SU(1,1)$ must have unit-modulus eigenvalues $e^{\pm i\theta}$, by the group definition. Consequently, in the unstable regime, $\mathcal{U}_F$ cannot be diagonalized by any symplectic similarity transformation $S\in SU(1,1)$ which preserves the bosonic commutation relation. 

Diagonalization in this regime requires a non-symplectic similarity transformation and therefore does not preserve the bosonic commutation relations. Although such a transformation is unphysical in the sense discussed above, we introduce the following relation for convenience:
\begin{equation}
    e^{\kappa_kT\sigma_x }=V\,\text{diag}(e^{\kappa_kT}, e^{-\kappa_kT})\,V^{-1},
\end{equation}
where $V=\dfrac{1}{\sqrt{2}}\begin{pmatrix}
    1 & 1 \\ 1 & -1
\end{pmatrix}\notin SU(1,1)$. Together with Eq.~(\ref{eq:hyp}), we know that there exists $S\in SU(1,1)$ such that 
\begin{equation}
\begin{aligned}
        \mathcal{U}_F&=S\exp[\kappa_k{T} \sigma_x-iT\mu\mathbb{I}]S^{-1}\\
                &=(SV)\,\begin{pmatrix}
                    e^{T\kappa_k-iT\mu} & 0 \\ 0 & e^{-T\kappa_k-iT\mu}
                \end{pmatrix}\,(SV)^{-1}
\end{aligned}
\end{equation}
Taking the logarithm yields the following result.
\begin{equation}
    \log\mathcal{U}_F=-iT\big[S(i\kappa_k \sigma_x+\mu\mathbb{I})S^{-1}\big].
    \label{eq:effdyn}
\end{equation}
Here we have used the relation $V\sigma_z V^{-1}=\sigma_x$. By a similar argument discussed in the previous case, we can let $\mathbf\Psi\equiv S^{-1}\mathbf\Phi=(b_k,\,b_{-k}^\dagger)^{\mathsf T}$, and derive the effective Hamiltonian $H_{\text{eff}}$. It turns out that the result is as follows:
\begin{equation}
    H_{\text{eff}}=i\kappa_k(b_k^\dagger b_{-k}^\dagger-b_{-k}b_{k})+\mu(b_k^\dagger b_k- b_{-k}^\dagger b_{-k})-\mu.
\end{equation}
Consequently, this is the regime where the system cannot be diagonalized into a stable harmonic oscillator form. Instead, Eq.~(\ref{eq:hyp}) identifies the canonical hyperbolic representative of $\mathcal{U}_F$, and Eq.~(\ref{eq:effdyn}) yields the corresponding effective dynamical matrix $i\kappa_k \sigma_x+\mu\mathbb{I}$ in the transformed basis $\mathbf\Psi=S^{-1}\mathbf\Phi$. Up to the identity shift $\mu \mathbb{I}$, the nontrivial part $i\kappa_k \sigma_x$ serves as the reference dynamical matrix in the unstable regime, as introduced in Sec.~F.

At this point, it is worth emphasizing that $ \epsilon_k $ and $\kappa_k$ are the quasienergy spectrum that we have obtained from the classical spectrum. This equivalence is a direct consequence of the classical-to-quantum correspondence established in Sec.~A, which ensures that both the classical and the quantum BdG operator dynamics are governed by the same dynamical matrix. Fig.~D.2(a) shows this shared spectrum, which is obtained from the monodromy matrix. Since it remains in the $SU(1,1)$ group, the eigenvalue pair of the monodromy matrix is given as 
\begin{equation}
  \lambda_+ \lambda_- = 1, \quad
  \lambda_\pm(k) =
  \begin{cases}
    e^{\mp i \epsilon_kT} \;&(k\notin\text{MG})\\
     e^{\pm\kappa_kT-iT\mu}     &(k\in \text{MG})
  \end{cases},
  \label{eq:SU11_eigs}
\end{equation}
where MG denotes the momentum-gap region. In our system, the MG region corresponds to the unstable regime, whereas the band region ($k\notin \text{MG}$) corresponds to the stable regime.
Interpreting the effective Hamiltonian $H_{\text{eff}}$, inside the band the system clearly has $\{\ket{n_k, m_{-k}}_b\}_{n,m}$ as its Floquet eigenstate at each mode, and has its corresponding quasienergy $(n_k+m_{-k}+1)\,\epsilon_k$. 
Because the original Hamiltonian is given as $H_k(t)=\bigoplus_qH_q(t)$, where integer $q$ indicates the number-difference sector, i.e., $q=n_k-n_{-k}$, we can treat each sector separately. As shown in Fig.~D.2(b), the spectrum collapses within each sector $q$ at the MG edge. This collapsing nature is inherited from the BdG spectrum, in which at the momentum gap edge, $\epsilon_k\rightarrow \Omega/2$.

\section{E. Petermann factor in quantum formalism}\label{smsec:E}
In this section, we investigate the role of the Petermann factor (PF) in the quantum regime. The PF was originally introduced in the context of classical non-Hermitian systems as a measure of mode non-orthogonality, and can be consistently defined for the classical monodromy matrix in Floquet systems. The PF $K_k$ defined for each mode has the form
\begin{equation}
        K_k = \frac{\braket{L_k|L_k}\braket{R_k|R_k}}
                {|\braket{L_k|R_k}|^2},
\end{equation}
where the right eigenvector $\ket{R_k}$ is defined via 
\begin{equation}
    \mathcal{U}_F\ket{R_k}=\lambda_k\ket{R_k},
\end{equation} 
and its corresponding left eigenvector $\bra{L_k}$ is defined as 
\begin{equation}
    \bra{L_k}\mathcal{U}_F=\bra{L_k}\lambda_k.
\end{equation}
This mode overlap is a classical quantity, but due to the correspondence discussed in Sec~A, it directly relates to the quantities in the quantum BdG matrix as well. We separate it into two cases: stable and unstable case to see how the PF manifests itself in each regime.

First, we consider the case in the stable regime, where the monodromy matrix can be diagonalized by a similarity transformation
\begin{equation}
    \mathcal U_F=S\,\begin{pmatrix}
        e^{- iT\epsilon_k} & 0 \\ 0&e^{ iT\epsilon_k} 
    \end{pmatrix}\,S^{-1},
\end{equation} 
with $S\in SU(1,1)$. Then, we can parameterize $S$ as 
\begin{equation}
    S=\begin{pmatrix}e^{-i\alpha}\cosh r_k & -e^{i\beta}\sinh r_k \\ -e^{-i\beta}\sinh r_k&e^{i\alpha}\cosh r_k\end{pmatrix},\quad
S^{-1}=\begin{pmatrix}e^{i\alpha}\cosh r_k & e^{i\beta}\sinh r_k \\ e^{-i\beta}\sinh r_k&e^{-i\alpha}\cosh r_k\end{pmatrix}, 
\label{eq:par}
\end{equation}
and it naturally implies that the Bogoliubov operator takes the form 
\begin{align}
    b_k=&e^{i\alpha}\cosh{r_k} \,a_k+e^{i\beta}\sinh{r_k}\,a^\dagger_{-k}\\
    b_{-k}^\dagger=&e^{-i\beta}\sinh{r_k} \,a_k+e^{-i\alpha}\cosh{r_k}\,a^\dagger_{-k}.
\end{align}
Here, the relation $SD=\mathcal{U}_F S$ directly gives the relation of the right eigenvector $v_i$, i.e.,  \begin{equation}
    \lambda_i v_i=\mathcal{U}_F \,v_i,\quad S=[v_1,\,v_2],
\end{equation} and the relation $DS^{-1}=S^{-1}\mathcal{U}_F$ gives the left eigenvector $w_i^\dagger$ defined by \begin{equation}
    w_i^\dagger \lambda_i = w_i^\dagger \,\mathcal{U}_F,\quad S^{-1}=\begin{bmatrix}
    w_1^\dagger\\w_2^\dagger
\end{bmatrix}.
\end{equation}
A straightforward calculation yields
\begin{align}
\braket{R^\pm|R^\pm}=&\cosh^2{r_k}+\sinh^2{r_k}=\cosh{2r_k}\\
\braket{L^\pm|L^\pm}=&\cosh{2r_k}\\
\braket{L^\pm|R^\pm}=&\cosh^2{r_k}-\sinh^2{r_k}= 1,
\end{align}
which gives the relation $K_k=\cosh^2{2r_k}$.

Meanwhile, the result can be recast in a different form as well. Within the same parameterization as in Eq.~(\ref{eq:par}), the effective dynamical matrix $\mathcal{M}_q^{\text{eff}}\equiv \sigma_zH_{\text{BdG}}^{\text{eff}}$ defined via
\begin{equation}
\begin{aligned}
    \mathcal{U}(T,0)=&\mathcal{T}\exp\bigl[-i\int_0^T \mathcal{M}_q(t)\,dt\bigr]\\
    =&\exp[-i \sigma_zH_{\text{BdG}}^{\text{eff}}T]  
\end{aligned}
\end{equation}
can be explicitly written as 
\begin{equation}
\begin{aligned}
    \mathcal{M}_q^{\text{eff}}
    =&\epsilon_k\begin{pmatrix}
        \cosh{2r_k} & e^{-i(\alpha-\beta)}\sinh{2r_k} 
        \\
        -e^{i(\alpha-\beta)}\sinh{2 r_k} & - \cosh{2r_k}
    \end{pmatrix}\\
    \equiv&\begin{pmatrix}
        g_k  & \Delta_k \\ -\Delta_k ^* & -g_k 
    \end{pmatrix},
\end{aligned}
\end{equation}
and therefore
\begin{equation}
    H_{\text{BdG}}^{\text{eff}}=\begin{pmatrix}
        g_k  & \Delta_k \\ \Delta_k ^* & g_k 
    \end{pmatrix}.
\end{equation}
Here, we parameterize the matrix elements by $g_k$ and $\Delta_k$, where the condition $g_k>|\Delta_k|$ follows directly from $\cosh{2 r_k}>\sinh{2r_k}$ and satisfies the relation $\epsilon_k=\sqrt{g_k^2-|\Delta_k|^2}$.
The PF can therefore be recast as:
\begin{equation}
    K_k=\frac{g_k^2}{g_k^2-|\Delta_k|^2}=\frac{g_k^2}{\epsilon_k^2}.
\end{equation}

Now, consider the case of an unstable regime. As mentioned in Sec.~D, the monodromy matrix in an unstable regime can be diagonalized as
\begin{equation}
   \mathcal{U}_F=-(SV)\text{diag}(e^{T \kappa_k},\,e^{-T\kappa_k})(SV)^{-1},
\end{equation} 
and the effective dynamical matrix is parametrized into 
\begin{align}
    \mathcal{M}_q^{\text{eff}}=&S(i \kappa_k \sigma_x+\mu\mathbb{I})S^{-1}
    \label{eq:effMat}
    \\
    \equiv&\begin{pmatrix}
        g_k  & \Delta_k \\ -\Delta_k ^* & -g_k
    \end{pmatrix} +\mu\mathbb{I},
\end{align}
with parameters $g_k$ and $\Delta_k$, satisfying $g_k<|\Delta_k|$. Since $S\in SU(1,1)$, parametrization analogous to Eq.~(\ref{eq:par}) can be applied to Eq.~(\ref{eq:effMat}), which gives 
\begin{equation}
    \begin{aligned}
        g_k=&\kappa_k\sin(\alpha+\beta)\sinh{2\eta_k} \\ 
        \Delta_k=& i\kappa_k \,e^{-2i\alpha}(\cosh^2{\eta_k}-e^{2i(\alpha+\beta)}\sinh^2{\eta_k}).
    \end{aligned}
\end{equation}
It directly leads to the relation $|\Delta_k|^2-g_k^2=\kappa_k^2$, where $\kappa_k$ denotes the imaginary part of the quasienergy spectrum.

Now, we derive an explicit form of the PF in an unstable regime. We start with the fact that the right eigenvectors $\ket{R^\pm_k}$ satisfy the relation
\begin{equation}
    \mathcal{U}_F\ket{R^\pm_k}=-e^{\pm T\kappa_k}\ket{R^\pm_k}.
\end{equation}
Based on this relation, we establish the relationship between the left and right eigenvectors. Omitting the mode index for brevity, the derivation proceeds as:
\begin{equation}
    \begin{aligned}
        (\bra{R^\pm}\sigma_z)\,\mathcal{U}_F=&\bra{R^\pm}(\mathcal{U}_F^\dagger)^{-1}\sigma_z\\
        =& (\mathcal{U}_F^{-1}\ket{R^\pm})^\dagger\sigma_z\\
        =&-e^{\mp T\kappa_k }\bra{R^\pm}\sigma_z
    \end{aligned}
\end{equation}
where the first line follows from the $\sigma_z$-pseudo-unitarity $\mathcal{U}_F^\dagger\,\sigma_z\,\mathcal{U}_F=\sigma_z$. In the final step, we used the fact that $\ket{R^\pm}$ having their eigenvalue $\lambda_\pm=-e^{\pm T\kappa_k}$ is an eigenvector of $\mathcal{U}_F^{-1}$ with the reciprocal eigenvalue $\lambda_\pm^{-1}=-e^{\mp T\kappa_k}$.
This confirms the relation that we can set the left eigenvector as $\bra{L^\pm}=\bra{R^\mp}\sigma_z$ in an unstable regime.

Then we can reformulate the form of the PF in an unstable regime as follows:
\begin{equation}
    K_k=\frac{\braket{R^\mp|R^\mp}\braket{R^\pm|R^\pm}}{|\braket{R^\mp|\sigma_z|R^\pm}|^2}.
\end{equation}
Rather than solving the eigenvalue problem with the monodromy matrix, we can work with the effective dynamical matrix, omitting the identity term $\mu\mathbb{I}$. That is, 
\begin{equation}
    \begin{pmatrix}
        g_k  & \Delta_k \\ -\Delta_k ^* & -g_k
    \end{pmatrix}\ket{R^\pm_k}=\pm i \kappa_k\,\ket{R^\pm_k}.
\end{equation}
Then, the right eigenvectors can be written as
\begin{equation}
    \ket{R_k^+}=\sqrt{N}\begin{pmatrix}
        \Delta_k \\ i\kappa_k-g_k
    \end{pmatrix},
    \qquad     \ket{R_k^-}=\sqrt{M}\begin{pmatrix}
        -\Delta_k \\ i\kappa_k+g_k
    \end{pmatrix},
\end{equation} with normalization constants $\sqrt{N}$ and $\sqrt{M}$.
Then, straightforward calculation yields
\begin{equation}
\begin{aligned}
    \braket{R^+|R^+}=&N\{|\Delta_k|^2+(g_k^2+\kappa_k^2)\}\\
=& 2N|\Delta_k|^2 
\end{aligned}
\end{equation}
\begin{equation}
\begin{aligned}
    \braket{R^-|R^-}=2M|\Delta_k|^2 
\end{aligned}
\end{equation}
\begin{equation}
\begin{aligned}
\braket{L^\pm|R^\pm}=&\braket{R^\pm|\sigma_z|R^\mp}\\
=&\sqrt{NM}\{(-|\Delta_k|^2+g_k^2-\kappa_k^2)\pm i(2g_k\kappa_k)\}\\
=& 2\sqrt{NM}\{-\kappa_k^2\pm ig_k\kappa_k\}.
\end{aligned}
\label{eq:pf1}
\end{equation}
Eq.~(\ref{eq:pf1}) gives $|\braket{R^\pm|\sigma_z|R^\mp}|^2=4NM\,\kappa_k^2 \,|\Delta_k|^2$, and therefore, satisfies the relation
\begin{equation}
    K_k=\frac{|\Delta_k|^2}{\kappa_k^2}.
\end{equation}

Consequently, we can express the Petermann factor in a unified form depending on the ratio between $g_k$ and $|\Delta_k|$:
\begin{equation}
   K_k= \begin{cases}
        \dfrac{1}{1-|\Delta_k|^2/g_k^2}&(k\notin\text{MG})\\ 
        \dfrac{1}{1-g_k^2/|\Delta_k|^2} &(k\in\text{MG}),
        
    \end{cases}
    \label{eq:PF}
\end{equation}
with $g_k^2/|\Delta_k|^2$ being a dimensionless parameter, where $|\Delta_k|^2<g_k^2$ inside the band (stable) and $|\Delta_k|^2>g_k^2$ in the momentum gap (unstable). Importantly, Eq.~(\ref{eq:PF}) is not specific to PTCs: it holds for any \textit{static} two-mode quadratic bosonic Hamiltonian whose BdG dynamical matrix can be written as $\sigma_z H_{\mathrm{BdG}}=\bigl(\begin{smallmatrix} g & \Delta \\ -\Delta^*&-g \end{smallmatrix}\bigr)$. In PTCs, the parameters $g_k$ and $\Delta_k$ are defined as the matrix elements of the \textit{effective} dynamical matrix $\sigma_z H_{\mathrm{BdG}}^\text{eff}$. Hence, the same logic leads to Eq.~(\ref{eq:PF}) both for stable ($|\Delta|<g$) and unstable ($|\Delta|>g$) regimes.

\section{F. Petermann factor as a squeezing measure in the unstable regime}\label{smsec:F}
In this section, we investigate how the Petermann factor manifests itself in the squeezing parameter in an unstable regime. Here, for the general discussion, we consider the static Hamiltonian given by
\begin{equation}
    H_0=g(n_k+n_{-k}+1)+(\Delta^* a_k a_{-k}+\text{h.c.}).
    \label{eq:sF1}
\end{equation}
Then the dynamical matrix that governs the Nambu operator dynamics, i.e., $ i\partial _t\mathbf\Phi=\sigma_z H_{\text{BdG}}\mathbf\Phi,\,\mathbf\Phi=(a_k ,\,a_{-k}^\dagger)^\mathsf{T}$ is given as follows:
\begin{equation}
    \sigma_zH_{\text{BdG}}=\begin{pmatrix}g & \Delta \\ -\Delta^* & -g
\end{pmatrix}.
\label{eq:sF2}
\end{equation}

First, we define the growth rate $\kappa=\sqrt{|\Delta|^2-g^2}$ in the unstable regime $|\Delta|>g$. Following Sec.~E, the Petermann factor reads
\begin{equation}
K=\frac{|\Delta|^2}{\kappa^2}=\dfrac{1}{1-g^2/|\Delta|^2}.
\end{equation}

Meanwhile, the evolution $\mathcal{U}(t,0)$ takes the form
\begin{equation}
    \begin{aligned}
        \mathcal{U}(t,0)=&\exp{(-it\sigma_zH_{\text{BdG}})}\\
        =& \cosh(\kappa t)\mathbb{I}-i\dfrac{\sinh(\kappa t)}{\kappa}(\sigma_zH_{\text{BdG}}).
    \end{aligned}
\end{equation}
From here, we obtain the relation of the two-mode squeezing parameter $r(t)=\text{arcsinh}(\frac{|\Delta|}{\kappa}\sinh(\kappa t))$, i.e.,
\begin{equation}
\begin{aligned}
r(t)=\text{arcsinh}(\sqrt{K}\sinh(\kappa t)).
\end{aligned}
\label{eq:sF3}
\end{equation}
To gain physical insight, we first consider the regime where $K\approx1$, i.e., $|\Delta|^2 \gg g^2$. Eq.~(\ref{eq:sF1}) and (\ref{eq:sF2}) directly gives that such a case is where $H_0\approx\Delta^* a_k a_{-k}+\text{h.c.}$. In this limit, the two-mode squeezing parameter grows linearly in time, i.e., $r(t)\approx\kappa t$.

Next, we consider the general case where $K>1$. In this regime, the previous approximation no longer holds, and the squeezing parameter obeys the relation given in Eq.~(\ref{eq:sF3}). 

In the short-time limit \(\sqrt{K}\kappa t\ll1\), the squeezing parameter reduces to \(r(t)\approx\sqrt{K}\kappa t\). Consequently, the initial growth rate is given by
\begin{equation}
    \frac{d r}{dt}\bigg|_{t\rightarrow0}\approx \sqrt{K}\kappa .
\end{equation}
Thus, for a fixed growth rate $\kappa$, the Petermann factor $K$ enhances the initial squeezing process, resulting in a growth rate that is $\sqrt{K}$ times larger than that of the case $K=1$.

Now, based on the above discussions, we further clarify the meaning of the PF in the unstable regime. For this, we introduce a reference dynamical matrix 
\begin{equation}
    \mathcal{M}_{\text{ref}}=\begin{pmatrix}
        0 & i\kappa \\ i\kappa & 0 
    \end{pmatrix},
\end{equation}
which represents a system with the growth rate $\kappa$, and Petermann Factor $K=1$. Up to a trivial identity shift $\mu\mathbb{I}$, any dynamical matrix in the unstable class with the same growth rate $\kappa$ can be related to this reference form $\mathcal{M}_{\text{ref}}=i\kappa\sigma_x$ via a similarity transformation $\sigma_zH_{\text{BdG}}=S\mathcal{M}_{\text{ref}}\,S^{-1}$ where $S\in SU(1,1)$~\cite{S-Inglima:2018oda}.
Within this framework, the PF can be interpreted as a prefactor that quantifies the deviation from the reference matrix $\mathcal{M}_{\text{ref}}$. While the growth rate $\kappa$ remains invariant under a transformation $S$, the degree of basis change induced by $S$ determines the magnitude of $K$, causing the initial growth rate to be amplified by a factor $\sqrt{K}$.

In the case of an unstable regime of PTCs, the effective dynamical matrix is given by Eq.~(\ref{eq:effdyn}). Consequently, in the Bogoliubov basis $\mathbf\Psi=S^{-1}\mathbf\Phi\equiv(b_k,\,b_{-k}^\dagger)^{\mathsf T}$ introduced in Sec.~D, the dynamical matrix is transformed into the reference form $i\kappa_k \sigma_x+\mu\mathbb{I}$ up to the identity shift $\mu\mathbb{I}$. Hence, the $b$-operator basis in Sec.~D can be viewed as a canonical \emph{reference squeezing basis} in which the transformed effective dynamical matrix takes the standard form $\mathcal{M}_{\mathrm{ref}}$.

\section{G. Effect of loss on the Petermann-squeezing relation}\label{smsec:G}
We now investigate the robustness of the Petermann-enhanced initial squeezing rate against intrinsic loss. Here we consider the Heisenberg-Langevin equations, which are given by
    \begin{equation}\begin{aligned}
    \partial_t{a_1} &= \bigg[ -ig - \frac{\gamma}{2} \bigg] a_1 - i\Delta a_2^\dagger + \sqrt{\gamma} a_{1,\text{in}}  \\
    \partial_t{a_2^\dagger} &= \bigg[ ig - \frac{\gamma}{2} \bigg] a_2^\dagger + i\Delta^* a_1 + \sqrt{\gamma} a_{2,\text{in}}^{\dagger}, 
\end{aligned}\end{equation}
assuming linear loss $\gamma_1=\gamma_2\equiv\gamma$ and coupling to a Markovian vacuum bath. Then the corresponding noise operators satisfy the relations
\begin{equation}
\begin{aligned}
    \braket{a_{\mu,\text{in}}(t)\,a^\dagger_{\nu,\text{in}}(t')}=&\delta_{\mu\nu}\delta(t-t'),\\
    \braket{a_{\mu,\text{in}}^\dagger(t)\,a_{\nu,\text{in}}(t')}=&0,
\\ \braket{a_{\mu,\text{in}}(t)\,a_{\nu,\text{in}}(t')}=&0.
\end{aligned}
\end{equation}

Casting the equations in the matrix form gives
\begin{equation}
    \dfrac{d}{dt}\begin{pmatrix}
        {a_1}\\
        {a_2^\dagger}
    \end{pmatrix}
    = -i\big[\mathcal{M}-i\frac{\gamma}{2}\mathbb{I}\big]\begin{pmatrix}
        a_1\\
        a_2^\dagger
    \end{pmatrix}+\sqrt{\gamma }\begin{pmatrix}
        a_{1,\text{in}}\\
        a_{2,\text{in}}^{\dagger}
    \end{pmatrix} .
\end{equation}
Solving the eigenvalue problem $\det\big(\mathcal{M}-(\lambda +i\dfrac{\gamma}{2})\mathbb{I}\big)=0$ yields $\lambda_{\pm}=\pm\sqrt{g^2-|\Delta|^2}-i\gamma/2$, and the threshold limit, marking the transition to the newly defined unstable regime (where $\exists \lambda \text{ s.t. }\mathfrak{Im}(\lambda)>0$) is determined by
\begin{equation}
    |\Delta|^2-g^2=\gamma^2/4.
\end{equation}

Now, it is natural to classify the parameter space into three distinct regimes: $\mathrm{(i)}\; g>|\Delta|$, corresponding to the elliptic regime discussed in Eq.~(\ref{eq:trace}), $\mathrm{(ii)}\;g^2<|\Delta|^2<g^2+\gamma^2/4$, representing the hyperbolic regime below the threshold, and $\mathrm{(iii)}\;|\Delta|^2>g^2+\gamma^2/4$, which corresponds to the regime above the threshold. Note that the exceptional point remains unchanged, but the stationary solution exists in the regimes $\mathrm{(i)}$ and $\mathrm{(ii)}$. 

Meanwhile, for symmetric linear loss, the matrix is shifted by $-\mathbb{I}\gamma/2$. Therefore the Petermann factor determined by the eigenvector biorthogonality remains unchanged, while the imaginary parts of the eigenvalues shift by $-\gamma/2$.

Assuming an initial vacuum state, we solve the dynamics of $\braket{a_1^\dagger a_1}=\braket{a_2^\dagger a_2}\equiv n$ and $\braket{a_1a_2}\equiv m$, governed by the coupled equations
\begin{equation}
    \begin{aligned}
        \frac{dn}{dt}=&-\gamma n+i\Delta^*m-i\Delta m^* \\
        \frac{dm}{dt}=&-(2ig+\gamma)m-i\Delta (2n+1).
    \end{aligned}
\end{equation}
Then the general solution is given by 
\begin{equation}
\begin{aligned}
        n(t)&=\frac{2|\Delta|^2}{\gamma^2-4(|\Delta|^2-g^2)}\bigg[1-e^{-\gamma t}\big(\cosh(2\kappa t)+\frac{\gamma}{2\kappa}\sinh(2\kappa t)\big)\bigg]\\
        |m(t)|&= \frac{1}{|\Delta|}\sqrt{(gn)^2+\frac{1}{4}(\gamma n+ \dot{n})^2},
\end{aligned}
\end{equation}
where $\sqrt{|\Delta|^2-g^2}\equiv \kappa$. Note that it is the general solution for all cases $\mathrm{(i)}-\mathrm{(iii)}$. In the regime where $|\Delta|^2<g^2$, substituting $\kappa=i\epsilon$ transforms the hyperbolic functions into trigonometric functions via the identities $\cosh(2i\epsilon t)=\cos( 2\epsilon t)$ and $\sinh(2 i\epsilon t)=i\sin(2\epsilon t)$.

Expanding $n(t)$ to the third order yields 
\begin{equation}
    n(t)=|\Delta|^2(t^2-\frac{2\gamma}{3}t^3)+\mathcal{O}(t^4),
\end{equation}
which shows that the loss $\gamma$ enters as a third-order correction.

Now, we derive the squeezing in the presence of loss, and investigate whether the relation between the initial growth rate and Petermann factor remains valid. We define the single-mode quadrature operator $X_\mu(\phi)=(a_\mu e^{-i\phi}+a^\dagger_\mu e^{i\phi})/\sqrt{2}$ for $\mu=1,2$.
Since the state remains Gaussian, it is natural to quantify the degree of squeezing by defining an effective squeezing parameter $r(t)$ via $ r(t)\equiv-\frac{1}{2}\ln{V_-(t)}$, where $V_-(t)$ denotes the minimum variance of the Einstein-Podolsky-Rosen (EPR) operator $X_-(\phi)\equiv X_1(\phi)-X_2(\phi)$. Note that this definition reduces to the squeezing parameter $r$ defined in the ideal unstable case in the previous section, when the loss goes to 0. 
Direct calculation yields
\begin{equation}
    V_-(t)=(2n+1)-2|m|,
\end{equation}
where we have used the relations $\braket{a_1^2}=\braket{a_2^2}=0$, and $\braket{a_1a^\dagger_2}=\braket{a_1^\dagger a_2}=0$.

Now, we can use the explicit value of $n(t)$ and $|m(t)|$ and examine the leading-order behavior of the squeezing parameter. 
Expanding to the first order (where $n(t)\sim\mathcal{O}(t^2)$ is negligible and $|m(t)|\approx|\Delta|t$), we obtain $V_-(t)\approx 1-2|\Delta|t$.
Relating it to the squeezing parameter in the regime $g<|\Delta|$ and using the relation $K=|\Delta|^2/\kappa^2$ gives 
\begin{equation}
    r(t)= \sqrt{K}\kappa t+\mathcal{O}(t^2),
\end{equation}
showing that for a fixed intrinsic growth rate $\kappa$, the enhancement factor $\sqrt{K}$ of the initial squeezing rate persists even in the presence of uniform loss. Therefore, the transient behavior at early time $t\ll\gamma^{-1}$ gives the desired relation between the initial squeezing rate and the Petermann factor.

Crucially, the relation remains valid even in regime $\mathrm{(ii)}$. Although this regime corresponds to an intrinsically unstable system that is stabilized by loss, the initial squeezing behavior is still determined by the Petermann factor, which is associated with the exceptional point $g=|\Delta|$.

\par\vspace{1em}
\noindent\textsuperscript{*}These authors contributed equally to this work.\par
\noindent\textsuperscript{$\dagger$}\texttt{kunx@cau.ac.kr}\par
\noindent\textsuperscript{$\ddagger$}\texttt{bmin@kaist.ac.kr}\par

\makeatletter
\patchcmd{\endthebibliography}
  {\label{LastBibItem}}
  {\label{LastBibItemSM}}
  {}{}
\makeatother

\phantomsection\label{smsec:refs}
\end{document}